\documentclass[sigconf]{acmart}
\AtBeginDocument{%
  }
\usepackage{makecell}
\usepackage{bm}
\usepackage{arydshln} 
\usepackage{setspace}
\usepackage{multirow}
\usepackage{hyperref}

\newcommand{\tabincell}[2]{\begin{tabular}{@{}#1@{}}#2\end{tabular}}
\copyrightyear{2022}
\acmYear{2022}
\setcopyright{rightsretained}
\acmConference[KDD '22] {Proceedings of the 28th ACM SIGKDD Conference on Knowledge Discovery and Data Mining}{August 14--18, 2022}{Washington, DC, USA}
\acmBooktitle{Proceedings of the 28th ACM SIGKDD Conference on Knowledge Discovery and Data Mining (KDD '22), August 14--18, 2022, Washington, DC, USA}
\acmISBN{978-1-4503-9385-0/22/08}
\acmDOI{10.1145/3534678.3539426}

\begin{document}

\title[KPGT]{KPGT: Knowledge-Guided Pre-training of Graph Transformer for Molecular Property Prediction}
\author{Han Li}
\email{lihan19@mails.tsinghua.edu.cn}
\affiliation{%
  \institution{Institute for Interdisciplinary Information Sciences, Tsinghua University}
  \streetaddress{30 Shuangqing Rd}
  \city{Haidian Qu}
  \state{Beijing Shi}
  \country{China}}
\author{Dan Zhao}
\authornotemark[1]
\email{zhaodan2018@tsinghua.edu.cn}
\affiliation{%
  \institution{Institute for Interdisciplinary Information Sciences, Tsinghua University}
  \streetaddress{30 Shuangqing Rd}
  \city{Haidian Qu}
  \state{Beijing Shi}
  \country{China}}
\author{Jianyang Zeng}
\authornote{Dan Zhao and Jianyang Zeng are the corresponding authors.}
\email{zengjy321@tsinghua.edu.cn}
\affiliation{%
  \institution{Institute for Interdisciplinary Information Sciences, Tsinghua University}
  \streetaddress{30 Shuangqing Rd}
  \city{Haidian Qu}
  \state{Beijing Shi}
  \country{China}}

\renewcommand{\shortauthors}{H. Li et al.}

\begin{abstract}
    Designing accurate deep learning models for molecular property prediction plays an increasingly essential role in drug and material discovery.
Recently, due to the scarcity of labeled molecules, self-supervised learning methods for learning generalizable and transferable representations of molecular graphs have attracted lots of attention. 
In this paper, we argue that there exist two major issues hindering current self-supervised learning methods from obtaining desired performance on molecular property prediction, that is, the ill-defined pre-training tasks and the limited model capacity.
To this end, we introduce Knowledge-guided Pre-training of Graph Transformer (KPGT), a novel self-supervised learning framework for molecular graph representation learning, to alleviate the aforementioned issues and improve the performance on the downstream molecular property prediction tasks.
More specifically, we first introduce a high-capacity model, named Line Graph Transformer (LiGhT), which emphasizes the importance of chemical bonds and is mainly designed to model the structural information of molecular graphs.
Then, a knowledge-guided pre-training strategy is proposed to exploit the additional knowledge of molecules to guide the model to capture the abundant structural and semantic information from large-scale unlabeled molecular graphs.
Extensive computational tests demonstrated that KPGT can offer superior performance over current state-of-the-art methods on several molecular property prediction tasks.

    \label{sections:abstract}
\end{abstract}



\maketitle
\section{Introduction}
    Molecular property prediction is of great significance to design novel molecules with desired properties for drug and material discovery~\cite{wu2018moleculenet,tkatchenko2020machine}.
Representing molecules as graphs, where nodes correspond to the atoms and edges correspond to the chemical bonds, deep learning models, especially graph neural networks (GNNs), have been widely used for molecular graph representation learning~\cite{gilmer2017neural,xiong2019pushing,zhou2020graph,yang2019analyzing,DBLP:conf/nips/CorsoCBLV20,beaini2021directional}.
However, due to the limited amount of labeled data and the giant chemical space, such deep learning models trained with supervised learning strategies often perform poorly, especially on the prediction of out-of-distribution data samples~\cite{wu2018moleculenet,DBLP:conf/iclr/HuLGZLPL20}.

Following the significant success of self-supervised learning methods in the fields of natural language processing (NLP)~\cite{DBLP:conf/naacl/DevlinCLT19,floridi2020gpt} and computer vision (CV)~\cite{DBLP:conf/cvpr/He0WXG20,DBLP:journals/corr/abs-2111-06377}, many recent works have proposed to employ self-supervised learning strategies to pre-train GNNs by leveraging the large-scale unlabeled molecules and have achieved superior prediction performance in comparison with supervised learning methods on the downstream molecular property prediction tasks~\cite{DBLP:conf/iclr/HuLGZLPL20,DBLP:conf/nips/LiuDL19,DBLP:conf/nips/YouCSCWS20,DBLP:conf/icml/XuWNGT21,DBLP:conf/icml/YouCSW21, DBLP:conf/nips/RongBXX0HH20,wang2021molclr,stark20213d,liu2021pre}.
Despite such fruitful progress, the prediction performance is still far from ideal.
In this paper, we argue that current self-supervised learning methods on molecular graphs still encounter the following two main issues:

\textbf{The ill-defined pre-training tasks.} The performance of self-supervised learning methods crucially depends on the design of pre-training tasks. 
So far, the self-supervised learning methods on molecular graphs can be roughly divided into two categories, that is, generative and contrastive methods according to their design of pre-training tasks. 
The generative methods follow the masked-language models in the NLP field, e.g., BERT~\cite{DBLP:conf/naacl/DevlinCLT19}, through masking a portion of molecular graphs, e.g., edges, nodes or subgraphs, and then learning to retrieve the original graphs~\cite{DBLP:conf/iclr/HuLGZLPL20,DBLP:conf/nips/RongBXX0HH20}. 
On the other hand, following the pioneering works in CV, the contrastive methods on molecular graphs first generate graph augmentations through strategies like node replacing, node dropping and edge perturbation for molecular graphs, and then learn to match the augmented graphs with the corresponding original molecular graphs in the embedding space~\cite{DBLP:conf/nips/YouCSCWS20,DBLP:conf/icml/YouCSW21}. 
However, unlike the word masking of human languages and the augmentation of images (e.g., resizing and rotation), which do not impact the fundamental semantics of the raw inputs, a small modification of molecular graphs can greatly change the characteristics of the corresponding molecules (an illustrative example is shown in Figure~\ref{figure:modification}).
Therefore, current self-supervised learning methods on molecular graphs can only capture the structural similarity of graphs and the simple construction rules (e.g., valency rule) of molecules, but fail to induce the abundant semantics related to molecular properties from chemical structures which are potentially more important to the downstream learning tasks.

\textbf{The limited model capacity}. 
Due to the giant chemical space and the wide breadth of the molecular properties ranging from quantum mechanics, physical chemistry, biophysics to physiology~\cite{wu2018moleculenet}, a high-capacity model is required to capture sufficient information from enormous unlabeled molecules.
Meanwhile, the success of self-supervised learning methods in the NLP and CV domains is indispensable to the emerging backbone networks with increasing amounts of parameters~\cite{fan2021beyond,liu2021swin}. 
Especially, transformer-based models have been proven to predominantly yield excellent prediction performance in these fields.
However, applying the transformer-based architectures in the self-supervised learning on molecular graphs is underexplored.
Most of the previous defined self-supervised learning methods on molecular graphs mainly employ GNNs, e.g., Graph Isomorphism Network (GIN)~\cite{DBLP:conf/iclr/XuHLJ19}, as the backbone networks, which provide only limited model capacity and potentially fail to capture the wide range of the information required for the prediction of various properties for a diversity of 
molecules. 

To this end, we propose Knowledge-guided Pre-training of Graph Transformer (KPGT), a novel self-supervised learning framework, to alleviate the aforementioned issues and learn more generalizable, transferable and robust representations of molecular graphs to improve the performance of the downstream molecular property prediction tasks. 
First, we propose a high-capacity model, named Line Graph Transformer (LiGhT), which represents molecular graphs as line graphs to emphasize the importance of chemical bonds, and also introduces path encoding and distance encoding to accurately preserve the structural information of molecules.
Then, we design a knowledge-guided pre-training strategy based on a generative self-supervised learning scheme.
Instead of directly predicting the randomly masked nodes using the contextual information, our proposed strategy leverages the additional knowledge of molecules (i.e., molecular descriptors and fingerprints), which serves as the semantics lost in the masked graph to guide the prediction of the masked nodes, thus making the model capture the abundant structural and semantic information from large-scale unlabeled molecules.

We conducted extensive computational experiments to evaluate the performance of our proposed method. 
We first pre-trained KPGT on a large-scale unlabeled molecule dataset and then applied the pre-trained model to eleven downstream molecular property datasets. 
The test results demonstrated that KPGT achieved superior results in comparison with current state-of-the-art self-supervised learning methods and thus verified the effective design of KPGT. 

To summarize, our work makes the following three main contributions:
(1) We point out that current self-supervised learning methods on molecular graphs have two major issues, that is, the ill-defined pre-training tasks and the limited model capacity;
(2) we propose KPGT, a novel self-supervised learning framework, consisting of a novel graph transformer architecture, LiGhT, and a knowledge-guided pre-training strategy, to alleviate the current issues in the representation learning of molecular graphs and improve the performance on the downstream molecular property prediction tasks;
(3) we conduct extensive tests to demonstrate that our proposed method can offer superior performance over current state-of-the-art methods on several molecular property prediction tasks.
The source code and datasets to reproduce our computational test results are available at \url{https://github.com/lihan97/KPGT}.

    \label{sections:introduction}

\section{Related Work}
    \subsection{Molecular Representation Learning}
Many efforts have been devoted to improving molecular representation learning for accurate molecular property prediction. 
The early feature-based methods exploited fixed molecular representation, e.g., molecular descriptors and fingerprints, to represent molecules in a vector space~\cite{van2003admet,dong2018admetlab,butler2018machine}. 
This kind of method highly relied on complicated feature engineering to produce promising prediction performance.
Then, deep learning based models were introduced to yield more expressive molecular representations. 
In particular, with their great advantage in modeling graph-structured data, GNNs had been widely applied for molecular graph representation learning.
For example, Gilmer et al.~\cite{gilmer2017neural} introduced a message passing framework for molecular property prediction.
Then many works proposed to modify the message passing mechanism to improve the expressiveness of GNNs~\cite{xiong2019pushing,DBLP:conf/nips/CorsoCBLV20,beaini2021directional}.
Please refer to~\cite{wieder2020compact} for a compact review on molecular property prediction with GNNs.
The message passing operators employed in the GNNs aggregated only local information and failed to capture the long-range dependence in molecules.
Recently, graph transformers have emerged to model such long-range dependencies. 
The applications of transformer-based architectures to molecular graphs have remained limited mainly due to the difficulty of properly modeling the structural information of molecular graphs.
Therefore, many efforts have been devoted to accurately preserving the structural information of molecular graphs~\cite{DBLP:journals/corr/abs-1905-12712, DBLP:journals/corr/abs-2002-08264,kreuzer2021rethinking,mialon2021graphit,ying2021do}.

\subsection{Self-Supervised Learning on Molecular Graphs}
Due to the data scarcity and the giant chemical space, self-supervised learning on molecular graphs has emerged as a core direction recently. 
Current self-supervised learning methods on molecular graphs can be roughly categorized into generative and contrastive ones, where they differed in the design of the pre-training tasks. 
The generative methods followed the masked-language model in the NLP field through masking nodes, edges or subgraphs and learning to retrieve the original graphs~\cite{DBLP:conf/iclr/HuLGZLPL20,DBLP:conf/nips/RongBXX0HH20}.
On the other hand, the contrastive methods followed the pioneer works in the CV field.
For example, You et al.~\cite{DBLP:conf/nips/YouCSCWS20} designed a framework with four graph augmentation strategies, that is, node dropping, edge perturbation, attribute masking and subgraph generating.
You et al.~\cite{DBLP:conf/icml/YouCSW21} further improved this framework by automatically and adaptively selecting data augmentations.
In addition, Xu et al.~\cite{DBLP:conf/icml/XuWNGT21} proposed to learn the hierarchical prototypes upon graph embeddings to infer the global-semantic structure in molecular graphs.
Moreover, St{\"a}rk et al.~\cite{stark20213d} and Liu et al.~\cite{liu2021pre} exploited the 3D geometry of molecules to improve the pre-training of GNNs.

    \label{sections:related_work}
    
\section{Knowledge-Guided Pre-training of Graph Transformer (KGPT)}
    In this section, we elaborate on the design of our self-supervised learning framework, Knowledge-guided Pre-training of Graph Transformer (KPGT).
First, we introduce how we transform molecular graphs into line graphs.
Then we provide the detailed implementation of our proposed Line Graph Transformer (LiGhT) for encoding the molecular line graphs.
Finally, we introduce our knowledge-guided pre-training strategy.
\subsection{Line Graph Transformer (LiGhT)}
\begin{figure}[htpb]
    \centering
    \includegraphics[width=\linewidth]{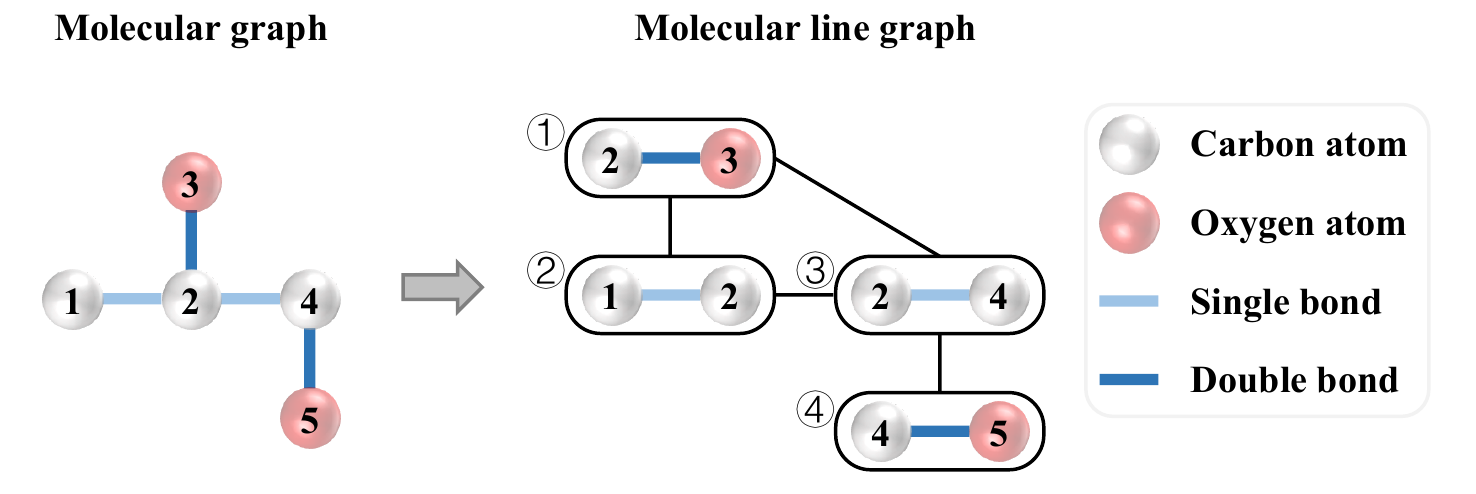}
    \doublespacing
    \caption{An illustrative example of the transformation of a molecular graph to a molecular line graph.}
    \label{figure:line_graph}
  \end{figure}
\subsubsection{Molecular Line Graph}
Given a molecule, it can be represented as a molecular graph $\mathcal{G}=(\mathcal{V},\mathcal{E})$, where $\mathcal{V}=\{v_i\}_{i\in[1,N_v]}$ stands for the set of nodes (i.e., atoms), $\mathcal{E}=\{e_{i,j}\}_{i,j\in[1,N_v]}$ stands for the set of edges (i.e., chemical bonds), and $N_v$ stands for the number of nodes. 
The initial features of node $v_i$ are represented by $\bm{x}^v_i\in\mathbb{R}^{D_v}$ and the initial features of edge $e_{i,j}$ are represented by $\bm{x}^e_{i,j}\in\mathbb{R}^{D_e}$, where $D_v$ and $D_e$ stand for the dimensions of features of nodes and edges, respectively.

As an illustrative example shown in Figure~\ref{figure:line_graph}, a molecular graph $\mathcal{G}$ can be transformed to a molecular line graph $\hat{\mathcal{G}}=\{\hat{\mathcal{V}}, {\hat{\mathcal{E}}\}}$ as follows:
(1) For each edge in $\mathcal{G}$, e.g., $e_{i,j}$, create a node $\hat{v}_{i,j}$ in $\hat{\mathcal{G}}$;
(2) for every two edges in $\mathcal{G}$ that have a node in common, create an edge between their corresponding nodes in $\hat{\mathcal{G}}$.
%

Then, for each node $\hat{v}_{i,j}$ in $\hat{\mathcal{G}}$, its initial feature embedding $\bm{h}_{\hat{v}_{i,j}} \in \mathbb{R}^{D_{\hat{v}}}$ is defined as:
\begin{equation}
    \bm{h}_{\hat{v}_{i,j}} = concat(\bm{W}_v \bm{x}^v_i + \bm{W}_v \bm{x}^v_j, \bm{W}_e \bm{x}^e_{i,j}),
\end{equation}
where $D_{\hat{v}}$ stands for the dimension of the feature embeddings of nodes in molecular line graphs, $\bm{W}_v \in \mathbb{R}^{\frac{D_{\hat{v}}}{2} \times D_v} $ and $\bm{W}_e \in \mathbb{R}^{\frac{D_{\hat{v}}}{2} \times D_e}$ stand for the trainable projection matrices, and $concat(\cdot)$ stands for the concatenation operator. For clarity, we denote the nodes in a molecular line graph as $\hat{\mathcal{V}}=\{\hat{v}_i\}_{i \in [1,N_{\hat{v}}]}$ in the following sections, where $N_{\hat{v}}$ stands for the number of nodes in the molecular line graph.
\subsubsection{Graph Transformer Architecture}
\begin{figure}[htpb]
    \centering
    \includegraphics[width=\linewidth]{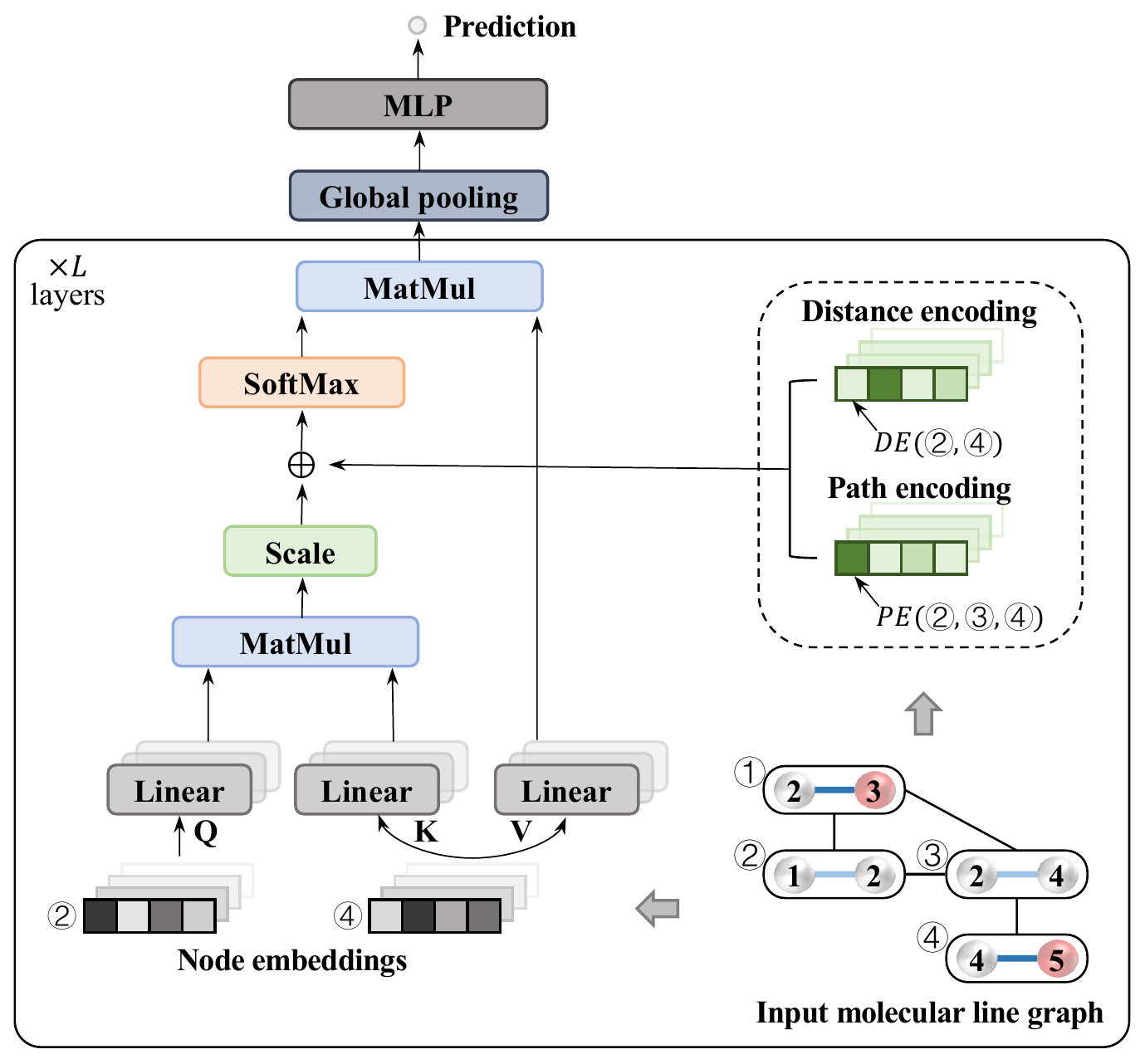}
    \caption{An illustration of the Line Graph Transformer (LiGhT) architecture. LiGhT is on the basis of a classic transformer encoder and introduces path encoding (PE) and distance encoding (DE) into the multi-head self-attention module to capture the structural information of molecules. }
    \label{figure:LiGhT}
  \end{figure}

In this section, we propose Line Graph Transformer (LiGhT), a novel transformer-based architecture, to encode the molecular line graphs and learn global feature representations for molecular property prediction (see Figure~\ref{figure:LiGhT}).
LiGhT is built upon a classic transformer encoder initially proposed in ~\cite{DBLP:conf/nips/VaswaniSPUJGKP17}.
The classic transformer consists of multiple transformer layers, each of which is composed of a multi-head self-attention module, followed by a feed-forward network (FFN) that includes residual connection and layer normalization operations.

More specifically, given a molecular line graph and the corresponding node feature matrix $\bm{H}\in \mathbb{R}^{N_{\hat{v}}\times D_{\hat{v}}}$, the multi-head self-attention module at layer $l$ is calculated as:
\begin{equation}
    \begin{aligned}
        \bm{Q}^{l,k} &= \bm{H}^{l-1} \bm{W}^{l,k}_Q, \bm{K}^{l,k} = \bm{H}^{l-1} \bm{W}^{l,k}_K,  \bm{V}^{l,k} = \bm{H}^{l-1} \bm{W}^{l,k}_V, \\
        \bm{A}^{l,k}&=softmax(\frac{\bm{Q}^{l,k} (\bm{K}^{l,k})^T}{\sqrt{D_h}}), \bm{H}^{l,k} = \bm{A}^{l,k}\bm{V}^{l,k}, \\
        \bm{H}^l &= concat(\bm{H}^{l,1},\bm{H}^{l,2},\dots,\bm{H}^{l,{N_h}}),
    \end{aligned}
    \label{eq:attention}
\end{equation}
where $\bm{H}^{l-1}$ stands for the node feature matrix at the $(l-1)$-th layer, $\bm{W}^{l,k}_Q \in \mathbb{R}^{D_{\hat{v}}\times D_h}$, $\bm{W}^{l,k}_K \in \mathbb{R}^{D_{\hat{v}}\times D_h}$ and $\bm{W}^{l,k}_V \in \mathbb{R}^{D_{\hat{v}}\times D_h}$ stand for the trainable projection matrices of the $k$-th head at layer $l$, $D_h=\frac{D_{\hat{v}}}{N_h}$ stands for the dimension of each self-attention head, $N_h$ stands for the number of self-attention heads, and $softmax(\cdot)$ stands for the softmax operator. 
The output $\bm{H}^l$ is then passed to a FFN:
\begin{equation}
    \begin{aligned}
       \hat{\bm{H}^l} &= LN(\bm{H}^{l-1}+\bm{H}^l),\\
       \bm{H}^l &= LN(\bm{W}^l_2GELU(\bm{W}^l_1\hat{\bm{H}^l})+\hat{\bm{H}^l}),
    \end{aligned}
\end{equation}
where $LN(\cdot)$ stands for the LayerNorm operator~\cite{DBLP:journals/corr/BaKH16}, $GELU(\cdot)$ stands for the GELU activation function~\cite{hendrycks2016gaussian}, and $\bm{W}^l_1\in \mathbb{R}^{4D_{\hat{v}}\times D_{\hat{v}}}$ and $\bm{W}^l_2\in \mathbb{R}^{D_{\hat{v}}\times 4D_{\hat{v}}}$ stand for the trainable projection matrices at layer $l$.

Unlike GNNs, which aggregate information from neighboring nodes, the global receptive field of the transformer enables each node to attend to the information at any position in one transformer layer.
Nevertheless, such a classic transformer architecture ignores the connectivity of graphs, thus causing a significant loss in the structural information of molecules.
In the NLP field, the structural information of chain-structured language can be preserved by giving each token a position embedding or encoding the relative distance between any two tokens~\cite{DBLP:conf/naacl/ShawUV18,DBLP:conf/nips/VaswaniSPUJGKP17}, which actually cannot be easily generalized to the graph-structured data.
Here, we introduce path encoding and distance encoding to inject the inductive bias in the multi-head self-attention module and thus enable LiGhT to encode the structural information of the molecular line graphs.

\textbf{Path Encoding.} 
For each pair of nodes $\hat{v}_i$ and $\hat{v}_j$ in the molecular line graph, we first derive the shortest path between them and then encode the path features to an attention scalar $a^p_{i,j}$ in a path attention matrix $\bm{A}^{p}\in \mathbb{R}^{N_{\hat{v}}\times N_{\hat{v}}}$ as follows:
\begin{equation}
    \begin{aligned}
    (\hat{v}^p_1, \hat{v}^p_2, \dots, \hat{v}^p_{N_p}) &= SP(\hat{v}_i, \hat{v}_j), \\
    a^p_{i,j} &= \bm{W}_a^p \frac{1}{N_p} \sum_{n=1}^{N_p} \bm{W}_n^p \bm{h}_{v^p_n},
    \end{aligned}
\end{equation}
where $SP(\cdot)$ stands for the shortest path function implemented by networkx~\cite{hagberg2008exploring}, $(\hat{v}^p_1, \hat{v}^p_2, \dots, \hat{v}^p_{N_p})$ stands for the shortest path between $\hat{v}_i$ and $\hat{v}_j$, $N_p$ stands for the length of the path, $\bm{h}_{v^p_n}$ stands for the feature of the $n$-th node in the shortest path, $\bm{W}_n^p \in \mathbb{R}^{D_p \times D_{\hat{v}}}$ stands for the trainable projection matrix for the $n$-th node in the path, $\bm{W}_a^p \in \mathbb{R}^{1 \times D_p}$ stands for a trainable projection matrix to project the path embedding to an attention scalar, and $D_p$ stands for the dimension of the path embedding.

\textbf{Distance Encoding.}
Following~\cite{DBLP:journals/corr/abs-2002-08264,ying2021do}, we also leverage the distances between pairs of nodes to further encode the spatial relations in the molecular line graphs.
More specifically, given nodes $\hat{v}_i$ and $\hat{v}_j$ in a molecular line graph, we encode their distance to an attention scalar $a^d_{i,j}$ in a distance attention matrix $\bm{A}^{d}\in \mathbb{R}^{N_{\hat{v}}\times N_{\hat{v}}}$ as follows:
\begin{equation}
    \begin{aligned}
        d_{i,j} &= SPD(\hat{v}_i,\hat{v}_j), \\
        a^d_{i,j} &= \bm{W}_2^d GELU(\bm{W}_1^d d_{i,j}),
    \end{aligned}
\end{equation}
where $SPD(\cdot)$ stands for the shortest path distance functoin, $d_{i,j}$ stands for the derived distance between $\hat{v}_i$ and $\hat{v}_j$, $\bm{W}_1^d \in \mathbb{R}^{D_d \times 1}$ and $\bm{W}_2^d \in \mathbb{R}^{1 \times D_d}$ stand for the trainable projection matrices, and $D_d$ stands for the dimension of the distance embedding.

Then, to introduce the encoded structural information into the model, we rewrite the formula of the attention matrix $\bm{A}^{l,k}\in \mathbb{R}^{N_{\hat{v}} \times N_{\hat{v}}}$ in the Eq.~\ref{eq:attention} as follows:
\begin{equation}
    \begin{aligned}
        \bm{A}^{l,k} = softmax(\frac{\bm{Q}^{l,k} (\bm{K}^{l,k})^T}{\sqrt{D_h}} + \bm{A}^p + \bm{A}^d).
    \end{aligned}
\end{equation}

Here, we discuss the main advantages of our proposed model compared with the previously defined graph transformers:

First, by representing molecular graphs as line graphs, LiGhT emphasizes the importance of chemical bonds in molecules.
Chemical bonds are the lasting attractions between atoms, which can be categorized into various types according to the ways they hold atoms together resulting in different properties of the formed molecules.  
However, the previously defined transformer architectures either omit the edge features or only introduce chemical bonds as the bias in the self-attention module, ignoring the rich information from chemical bonds~\cite{DBLP:journals/corr/abs-1905-12712, DBLP:journals/corr/abs-2002-08264,kreuzer2021rethinking,mialon2021graphit,ying2021do}. 
In our case, LiGhT fills this gap and fully exploits the intrinsic features of chemical bonds. 

Second, although strategies like path encoding have already been proposed in previous graph transformer architectures~\cite{DBLP:journals/corr/abs-1905-12712, ying2021do}, when encoding the paths, they only consider the edge features and ignore the node features in the paths.
On the other hand, our path encoding strategy incorporates the features of the complete paths between pairs of nodes, thus encoding the structural information more precisely compared to the previous methods.

In summary, LiGhT provides a reliable backbone network for accurately modeling the structural and semantic information of molecular line graphs.

\subsection{Knowledge-Guided Pre-training Strategy}
\begin{figure*}[htpb]
    \centering
    \includegraphics[width=0.8\linewidth]{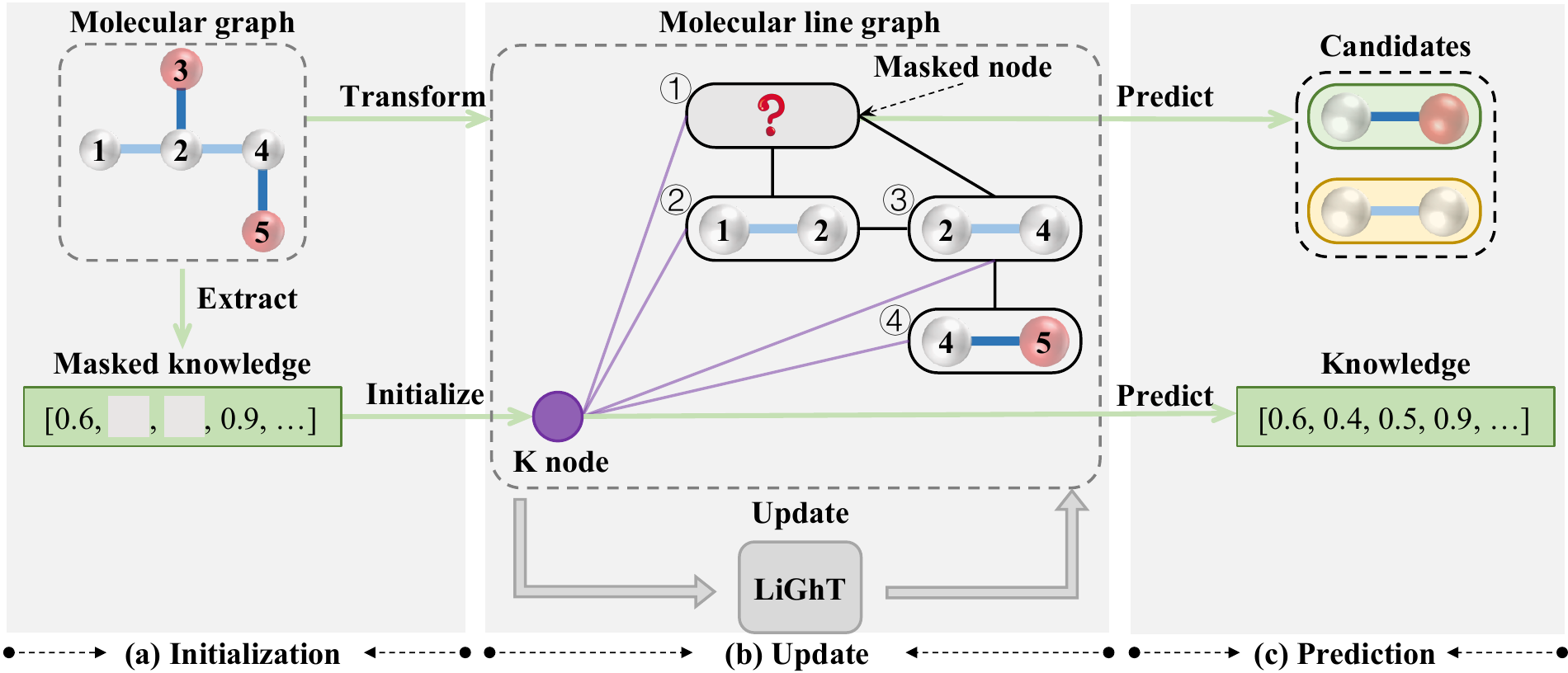}
    \caption{An illustration of the knowledge-guided pre-training strategy. 
    For a molecular graph, we first transform it to a line graph and extract its additional knowledge. 
Then, we randomly mask a proportion of the nodes in the line graph and the same proportion of the additional knowledge and utilize the masked knowledge to initialize the virtually constructed K node.
Next, LiGhT is employed to update the feature embeddings of nodes in the graph.
Finally, the derived node embeddings of the masked nodes and K node are used to predict the node types and masked knowledge, respectively.
    }
    \label{figure:KGP}
\end{figure*}
%
Our pre-training strategy is derived mainly based on a generative self-supervised learning scheme.
More specifically, we randomly choose a proportion of nodes in molecular line graphs for prediction. 
Following~\cite{DBLP:conf/naacl/DevlinCLT19}, for a chosen node, we first replace it with (1) a mask token $80\%$ of the time, (2) a random node 10\% of the time, and (3) the unchanged node 10\% of the time, and then predict the type of the original node with a cross-entropy loss.

As mentioned above, directly applying such masking operations may lose the initial semantics of molecular graphs, thus making current generative methods ill-defined.
In addition, without semantics, the only dependence between the masked nodes and their adjacent nodes is the valency rule, which is much weaker in comparison with the dependence between neighboring tokens in natural languages or neighboring pixels in images.
As an example shown in Figure~\ref{figure:mask}, if we mask an oxygen atom in the molecular graph, the oxygen atom is the expected prediction but many other atoms can also construct valid molecules according to the valency rules.
Therefore, such weak dependence cannot always guide the model to make the right predictions, which may cause the model to simply memorize the whole dataset.

To alleviate these issues, we introduce additional knowledge of molecules in our pre-training framework, which serves as the semantics lost in the masked graphs to guide the prediction of the masked nodes.

For molecules, there is a large number of well-defined molecular descriptors and fingerprints that have been widely used to represent molecular characteristics and have been proven to be effective for molecular property prediction in the previous research~\cite{van2003admet,dong2018admetlab,butler2018machine}. 
Molecular descriptors are typically generated by logical and mathematical procedures to quantitatively describe the physical and chemical profiles of molecules~\cite{xue2000molecular}. 
For example, LogP is a type of molecular descriptor measuring the lipophilicity of the molecules. 
The molecular fingerprints describe a molecular structure by a bit string, where one indicates the existence of one substructure and vice versa~\cite{rogers2010extended,cereto2015molecular}.
Therefore, molecular descriptors and fingerprints can provide valuable global and local information of molecules.
Moreover, most of the molecular descriptors and fingerprints are readily available by current cheminformatics tools~\cite{greg_landrum_2021_5589557, moriwaki2018mordred,Ramsundar-et-al-2019}. Thus, it will not introduce extra budget in the pre-training process. 

To incorporate such additional knowledge in our pre-training framework, we define a special node for each molecular line graph, named knowledge node (K node), and make connections between K node and each node individually. 
The raw features of the K node are initialized by the quantitative molecular descriptors and fingerprints.
As such, in the pre-training, the extra knowledge can be attended by other nodes through the self-attention module in the transformer and thus guide the prediction of the masked nodes.

In the pre-training, we also randomly mask a proportion of the initial features of K nodes and learn to predict the masked molecular descriptors and fingerprints. 
The prediction of the masked molecular descriptors is formulated as a regression task equipped with an RMSE loss, while the prediction of fingerprints is formulated as a binary classification task equipped with a cross-entropy loss. 

An illustrative example of our self-supervised learning strategy is shown in Figure~\ref{figure:KGP}. 
In summary, by explicitly introducing additional knowledge in the pre-training, KPGT is able to correct the current ill-defined generative pre-training tasks and thus enables the model to capture both the abundant structural and the rich semantics information from the molecular graphs, which can be essential to the downstream molecular property prediction tasks.

    \label{sections:method}
    
\section{Experiments}
    \begin{table*}[htpb]
  \caption{
      The AUROC performance of different methods on the classification datasets under both feature extraction and transfer learning settings. 
      The higher result is better (marked by $\uparrow$).
      AVG represents the averaging results over all the datasets. 
      The numbers in brackets are the standard deviations.
      The best result for each dataset is marked in bold and the second-best result is underlined.
    }
  \label{table:clf}
  \centering
\centering
\begin{tabular}{c|ccccccccc}
\hline
\multicolumn{1}{c|}{\multirow{3}{*}{Method}} & \multicolumn{9}{c}{Classification dataset $\uparrow$}                                                                    \\ \cline{2-10} 
\multicolumn{1}{c|}{}                        & \multicolumn{9}{c}{Feature extraction setting}                                                                    \\ \cline{2-10} 
\multicolumn{1}{c|}{}                        & BACE           & BBBP           & ClinTox        & SIDER & Estrogen & MetStab & Tox21          & ToxCast  & \textbf{AVG}  \\ \hline
Infomax~\cite{DBLP:conf/iclr/VelickovicFHLBH19}    & $0.825_{(0.033)}$ & $0.833_{(0.017)}$ & $0.637_{(0.027)}$ & $0.573_{(0.016)}$ & $0.878_{(0.061)}$ & $0.763_{(0.044)}$ & $0.764_{(0.038)}$ & $0.662_{(0.010)}$ & 0.742 \\
Edgepred~\cite{DBLP:conf/nips/HamiltonYL17}        & $0.807_{(0.014)}$ & $0.823_{(0.011)}$ & $0.637_{(0.059)}$ & $0.545_{(0.019)}$ & $0.851_{(0.070)}$ & $0.792_{(0.064)}$ & $0.702_{(0.040)}$ & $0.620_{(0.004)}$ & 0.722 \\
Masking~\cite{DBLP:conf/iclr/HuLGZLPL20}           & $0.739_{(0.059)}$ & $0.823_{(0.031)}$ & $0.564_{(0.089)}$ & $0.555_{(0.032)}$ & $0.805_{(0.075)}$ & $0.588_{(0.102)}$ & $0.722_{(0.017)}$ & $0.591_{(0.023)}$ & 0.673 \\
Contextpred~\cite{DBLP:conf/iclr/HuLGZLPL20}       & $0.822_{(0.053)}$ & $0.875_{(0.009)}$ & $0.607_{(0.062)}$ & $0.585_{(0.025)}$ & $0.872_{(0.059)}$ & $0.791_{(0.078)}$ & $0.757_{(0.028)}$ & $0.671_{(0.009)}$ & 0.748 \\
Masking+Sup~\cite{DBLP:conf/iclr/HuLGZLPL20}       & $0.818_{(0.039)}$ & $0.849_{(0.024)}$ & $0.792_{(0.043)}$ & $0.611_{(0.003)}$ & $\underline{0.888}_{(0.038)}$ & $\underline{0.819}_{(0.054)}$ & $0.828_{(0.013)}$ & $0.685_{(0.013)}$ & 0.786 \\
Contextpred+Sup~\cite{DBLP:conf/iclr/HuLGZLPL20}   & $0.831_{(0.018)}$ & $0.852_{(0.039)}$ & $0.765_{(0.061)}$ & $\underline{0.612}_{(0.026)}$ & $0.871_{(0.051)}$ & $0.807_{(0.039)}$ & $\underline{0.829}_{(0.026)}$ & $0.687_{(0.008)}$ & 0.782 \\
GraphLoG~\cite{DBLP:conf/icml/XuWNGT21}            & $0.766_{(0.040)}$ & $0.799_{(0.025)}$ & $0.526_{(0.101)}$ & $0.583_{(0.016)}$ & $0.867_{(0.071)}$ & $0.747_{(0.063)}$ & $0.691_{(0.041)}$ & $0.608_{(0.011)}$ & 0.698 \\
GraphCL~\cite{DBLP:conf/nips/YouCSCWS20}           & $0.802_{(0.045)}$ & $0.836_{(0.021)}$ & $0.663_{(0.018)}$ & $0.576_{(0.011)}$ & $0.864_{(0.050)}$ & $0.779_{(0.062)}$ & $0.758_{(0.036)}$ & $0.639_{(0.012)}$ & 0.740 \\
JOAO~\cite{DBLP:conf/icml/YouCSW21}                & $\underline{0.834}_{(0.033)}$ & $0.854_{(0.023)}$ & $0.747_{(0.039)}$ & $0.612_{(0.007)}$ & $0.854_{(0.061)}$ & $0.800_{(0.062)}$ & $0.801_{(0.023)}$ & $0.653_{(0.014)}$ & 0.769 \\
GROVER~\cite{DBLP:conf/nips/RongBXX0HH20}          & $0.809_{(0.046)}$ & $\underline{0.885}_{(0.009)}$ & $\underline{0.809}_{(0.048)}$ & $0.591_{(0.030)}$ & $0.870_{(0.048)}$ & $0.818_{(0.056)}$ & $0.827_{(0.019)}$ & $\underline{0.708}_{(0.015)}$ & \underline{0.790} \\
3DInfomax~\cite{stark20213d}                       & $0.810_{(0.044)}$ & $0.806_{(0.037)}$ & $0.743_{(0.014)}$ & $0.585_{(0.021)}$ & $0.852_{(0.054)}$ & $0.784_{(0.072)}$ & $0.738_{(0.021)}$ & $0.609_{(0.021)}$ & 0.741 \\
GraphMVP~\cite{liu2021pre}                         & $0.741_{(0.012)}$ & $0.836_{(0.028)}$ & $0.696_{(0.051)}$ & $0.563_{(0.023)}$ & $0.833_{(0.071)}$ & $0.669_{(0.057)}$ & $0.778_{(0.022)}$ & $0.645_{(0.006)}$ & 0.720 \\
KPGT                                               & $\textbf{0.868}_{(0.011)}$ & $\textbf{0.896}_{(0.010)}$ & $\textbf{0.856}_{(0.028)}$ & $\textbf{0.644}_{(0.018)}$ & $\textbf{0.890}_{(0.048)}$ & $\textbf{0.885}_{(0.047)}$ & $\textbf{0.838}_{(0.020)}$ & $\textbf{0.727}_{(0.016)}$ & \textbf{0.825} \\ \hline
\end{tabular}
\begin{tabular}{c|ccccccccc}
  \hline
  \multicolumn{1}{c|}{\multirow{2}{*}{Method}} & \multicolumn{9}{c}{Classification dataset $\uparrow$}                                                                    \\ \cline{2-10} 
  \multicolumn{1}{c|}{}                        & \multicolumn{9}{c}{Transfer learning setting}                                                                    \\ \cline{2-10} 
  \multicolumn{1}{c|}{}                        & BACE           & BBBP           & ClinTox        & SIDER & Estrogen          & MetStab & Tox21 & ToxCast & \textbf{AVG}  \\ \hline
  Infomax~\cite{DBLP:conf/iclr/VelickovicFHLBH19}        & $0.839_{(0.008)}$ & $0.840_{(0.026)}$ & $0.661_{(0.026)}$ & $0.616_{(0.024)}$ & $0.888_{(0.056)}$ & $0.837_{(0.066)}$ & $0.816_{(0.021)}$ & $0.690_{(0.012)}$ & 0.773 \\
  Edgepred~\cite{DBLP:conf/nips/HamiltonYL17}            & $0.817_{(0.034)}$ & $0.873_{(0.016)}$ & $0.730_{(0.017)}$ & $0.603_{(0.025)}$ & $0.881_{(0.061)}$ & $0.844_{(0.054)}$ & $0.818_{(0.025)}$ & $0.712_{(0.011)}$ & 0.785 \\
  Masking~\cite{DBLP:conf/iclr/HuLGZLPL20}               & $0.823_{(0.004)}$ & $0.864_{(0.028)}$ & $0.729_{(0.039)}$ & $0.573_{(0.012)}$ & $0.869_{(0.050)}$ & $0.868_{(0.061)}$ & $0.798_{(0.025)}$ & $0.663_{(0.018)}$ & 0.773 \\
  Contextpred~\cite{DBLP:conf/iclr/HuLGZLPL20}           & $0.840_{(0.009)}$ & $0.877_{(0.026)}$ & $0.732_{(0.017)}$ & $0.609_{(0.013)}$ & $0.882_{(0.065)}$ & $0.857_{(0.053)}$ & $0.806_{(0.012)}$ & $0.714_{(0.018)}$ & 0.790 \\
  Masking+Sup~\cite{DBLP:conf/iclr/HuLGZLPL20}           & $0.824_{(0.021)}$ & $0.859_{(0.020)}$ & $0.796_{(0.079)}$ & $0.606_{(0.005)}$ & $0.888_{(0.041)}$ & $0.872_{(0.051)}$ & $0.827_{(0.021)}$ & $0.715_{(0.007)}$ & 0.799 \\
  Contextpred+Sup~\cite{DBLP:conf/iclr/HuLGZLPL20}       & $\underline{0.855}_{(0.023)}$ & $0.875_{(0.011)}$ & $0.802_{(0.079)}$ & $0.620_{(0.009)}$ & $0.885_{(0.053)}$ & $0.859_{(0.055)}$ & $\underline{0.840}_{(0.023)}$ & $\underline{0.724}_{(0.015)}$ & 0.807 \\
  GraphLoG~\cite{DBLP:conf/icml/XuWNGT21}                & $0.830_{(0.014)}$ & $0.846_{(0.008)}$ & $0.667_{(0.021)}$ & $0.615_{(0.013)}$ & $0.871_{(0.054)}$ & $0.850_{(0.080)}$ & $0.796_{(0.025)}$ & $0.677_{(0.019)}$ & 0.769 \\
  GraphCL~\cite{DBLP:conf/nips/YouCSCWS20}               & $0.825_{(0.018)}$ & $0.887_{(0.019)}$ & $0.691_{(0.065)}$ & $0.587_{(0.026)}$ & $0.875_{(0.048)}$ & $0.821_{(0.066)}$ & $0.805_{(0.017)}$ & $0.696_{(0.023)}$ & 0.774 \\
  JOAO~\cite{DBLP:conf/icml/YouCSW21}                    & $0.826_{(0.029)}$ & $0.879_{(0.020)}$ & $0.741_{(0.047)}$ & $\underline{0.640}_{(0.010)}$ & $0.861_{(0.066)}$ & $0.837_{(0.058)}$ & $0.823_{(0.022)}$ & $0.711_{(0.014)}$ & 0.790 \\
  GROVER~\cite{DBLP:conf/nips/RongBXX0HH20}              & $0.840_{(0.030)}$ & $\underline{0.887}_{(0.006)}$ & $0.874_{(0.048)}$ & $0.638_{(0.005)}$ & $\underline{0.892}_{(0.044)}$ & $\underline{0.876}_{(0.038)}$ & $0.838_{(0.017)}$ & $0.696_{(0.014)}$ & \underline{0.818} \\
  3DInfomax~\cite{stark20213d}                           & $0.811_{(0.048)}$ & $0.877_{(0.014)}$ & $\underline{0.887}_{(0.033)}$ & $0.585_{(0.017)}$ & $0.880_{(0.054)}$ & $0.866_{(0.047)}$ & $0.805_{(0.032)}$ & $0.716_{(0.013)}$ & 0.804 \\
  GraphMVP~\cite{liu2021pre}                             & $0.818_{(0.012)}$ & $0.860_{(0.034)}$ & $0.719_{(0.044)}$ & $0.584_{(0.026)}$ & $0.865_{(0.061)}$ & $0.820_{(0.066)}$ & $0.799_{(0.018)}$ & $0.689_{(0.010)}$ & 0.769 \\
  KPGT                                          & $\textbf{0.855}_{(0.011)}$ & $\textbf{0.908}_{(0.010)}$ & $\textbf{0.946}_{(0.022)}$ & $\textbf{0.649}_{(0.009)}$  & $\textbf{0.905}_{(0.028)}$ & $\textbf{0.889}_{(0.047)}$ & $\textbf{0.848}_{(0.013)}$ & $\textbf{0.746}_{(0.002)}$ & \textbf{0.843}  \\ \hline
  \end{tabular}
\end{table*}
\subsection{Experimental Settings}
\textbf{Dataset.}
For pre-training, we used two million unique unlabeled molecular SMILES samples from the ChEMBL29 dataset~\cite{gaulton2017chembl}. 
We employed RDKit~\cite{greg_landrum_2021_5589557} to abstract the molecular graphs from the SMILES strings and initialize the features of nodes and edges (see Appendix~\ref{appendix:molecule} for more details).

After pre-training, we evaluated the effectiveness of KPGT on eleven molecular property datasets that have been widely used in the previous research, including eight classification datasets and three regression datasets, covering properties from multiple domains
(see Appendix~\ref{appendix:dataset} for more details about these datasets).

In practice, the test molecules can be structurally different from the training molecules.
Therefore, we used scaffold splitting, which split a dataset according to the graph substructures of the molecules in the dataset, offering a more challenging yet realistic scenario. 
We split each dataset into training, validation and test sets with a ratio of 8:1:1. 
Following the previous research~\cite{wu2018moleculenet,DBLP:conf/nips/RongBXX0HH20}, for each dataset, we performed three repeated tests with three randomly seeded splittings and then reported the means and standard deviations.

\textbf{Additional Knowledge.}
In this work, we did not carefully collect the additional knowledge, though it would better benefit KPGT, intuitively.
More specifically, we collected two sets of knowledge derived from RDKit~\cite{greg_landrum_2021_5589557}.
The first set contained 200 molecular descriptors that had been widely used for molecular property prediction~\cite{yang2019analyzing,DBLP:conf/nips/RongBXX0HH20}.
The second set contained the most common 512 RDKit fingerprints, where each bit indicated the existence of a specific path of length between one and seven in the molecules.

\textbf{Training Details.}
We mainly used Pytorch~\cite{paszke2017automatic} and DGL~\cite{wang2019dgl} to implement KPGT. 
More specifically, we implemented a 12-layer LiGhT as the backbone network with a hidden size of 768 and the number of self-attention heads as 12. 
A mean pooling operation that averaged all the nodes in individual graphs was applied on top of the model to extract molecular representations. 
An Adam optimizer~\cite{DBLP:journals/corr/KingmaB14} with weight decay $1e^{-6}$ and learning rate $2e^{-4}$ was used to optimize the model. 
The model was trained with batch size 1024 for a total of 100,000 steps. 
The KPGT had around 100 million parameters. 
We set the masking rate of both nodes and additional knowledge to 0.5.
The pre-training of KPGT took about two days on four Nvidia A100 GPUs.

For the downstream fine-tuning tasks, we added a predictor, which was a 2-layer multi-layer perceptron (MLP) with ReLU activation~\cite{DBLP:journals/corr/abs-1803-08375}, on top of the base feature extractors. 
The cross-entropy loss and RMSE loss were implemented for binary classification tasks and regression tasks, respectively. 
The model was trained using an Adam optimizer with batch size 32 for another 50 epochs. 
Early stopping was employed with patience set to 20.
For each method, we tried 48 hyper-parameter combinations to find the best results for each task.
For more training details please refer to Appendix~\ref{appendix:train}.

\textbf{Evaluation Protocols.}
The downstream molecular property prediction tests were performed under two evaluation protocols, that is, the feature extraction setting and the transfer learning setting.
For the feature extraction setting, we first fixed the pre-trained model and used it as a feature extractor to obtain the molecular graph representations of data samples, and then trained the predictor to make predictions.
For the transfer learning setting, we fine-tuned all the parameters in the model.

\textbf{Baselines.}
We comprehensively evaluated KPGT against twelve state-of-the-art self-supervised learning methods on molecular graphs, including six generative methods, i.e., Edgepred~\cite{DBLP:conf/nips/HamiltonYL17}, Masking~\cite{DBLP:conf/iclr/HuLGZLPL20}, Contextpred~\cite{DBLP:conf/iclr/HuLGZLPL20}, Masking+Sup~\cite{DBLP:conf/iclr/HuLGZLPL20}, Contextpred+Sup~\cite{DBLP:conf/iclr/HuLGZLPL20} and GROVER~\cite{DBLP:conf/nips/RongBXX0HH20}, and six contrastive methods, including Infomax~\cite{DBLP:conf/iclr/VelickovicFHLBH19}, GraphLoG~\cite{DBLP:conf/icml/XuWNGT21}, GraphCL~\cite{DBLP:conf/nips/YouCSCWS20}, JOAO\cite{DBLP:conf/icml/YouCSW21}, 3DInfomax~\cite{stark20213d} and GraphMVP~\cite{liu2021pre}. 

\subsection{Results on Downstream Tasks}
We reported the evaluation results of molecular property prediction under both feature extraction and transfer learning settings in Tables~\ref{table:clf} and~\ref{table:reg}. 
\begin{table*}
  \caption{
    The RMSE performance of different methods on the regression datasets under both feature extraction and transfer learning settings. 
    The lower result is better (marked by $\downarrow$).
    AVG represents the averaging results over all the datasets. 
    The numbers in brackets are the standard deviations.
    The best result for each dataset is marked in bold and the second-best result is underlined.
  }
  \label{table:reg}
  \centering
\centering
\begin{tabular}{c|cccccccc}
  \hline
  \multirow{3}{*}{Method} & \multicolumn{8}{c}{Regression dataset $\downarrow$}                                                                                       \\ \cline{2-9} 
                          & \multicolumn{4}{c|}{Feature extraction setting}                               & \multicolumn{4}{c}{Transfer learning setting} \\ \cline{2-9} 
                          & FreeSolv       & ESOL           & Lipo           & \multicolumn{1}{c|}{\textbf{AVG}}  & FreeSolv  & ESOL            & Lipo   & \textbf{AVG}   \\ \hline
  Infomax~\cite{DBLP:conf/iclr/VelickovicFHLBH19}                 & $4.119_{(0.974)}$ & $1.462_{(0.076)}$ & $0.978_{(0.076)}$ & \multicolumn{1}{c|}{2.186} & $3.416_{(0.928)}$ & $1.096_{(0.116)}$ & $0.799_{(0.047)}$ & 1.770 \\
  Edgepred~\cite{DBLP:conf/nips/HamiltonYL17}                & $3.849_{(0.950)}$ & $2.272_{(0.213)}$ & $1.030_{(0.024)}$ & \multicolumn{1}{c|}{2.384} & $3.076_{(0.585)}$ & $1.228_{(0.073)}$ & $0.719_{(0.013)}$ & 1.674 \\
  Masking~\cite{DBLP:conf/iclr/HuLGZLPL20}                 & $3.646_{(0.947)}$ & $2.100_{(0.040)}$ & $1.063_{(0.028)}$ & \multicolumn{1}{c|}{2.270} & $3.040_{(0.334)}$ & $1.326_{(0.115)}$ & $0.724_{(0.012)}$ & 1.697 \\
  Contextpred~\cite{DBLP:conf/iclr/HuLGZLPL20}             & $3.141_{(0.905)}$ & $1.349_{(0.069)}$ & $0.969_{(0.076)}$ &  \multicolumn{1}{c|}{1.820} & $2.890_{(1.077)}$ & $1.077_{(0.029)}$ & $0.722_{(0.034)}$ & 1.563 \\
  Masking+Sup~\cite{DBLP:conf/iclr/HuLGZLPL20}             & $3.210_{(0.876)}$ & $1.387_{(0.007)}$ & $\underline{0.725}_{(0.033)}$ & \multicolumn{1}{c|}{1.774} & $2.883_{(0.559)}$ & $1.297_{(0.114)}$ & $0.681_{(0.026)}$ & 1.620 \\
  Contextpred+Sup~\cite{DBLP:conf/iclr/HuLGZLPL20}         & $3.105_{(0.701)}$ & $1.477_{(0.038)}$ & $0.754_{(0.032)}$ & \multicolumn{1}{c|}{1.779} & $\underline{2.383}_{(0.624)}$ & $1.178_{(0.026)}$ & $0.681_{(0.030)}$ & 1.414 \\
  GraphLoG~\cite{DBLP:conf/icml/XuWNGT21}                & $4.174_{(1.077)}$ & $2.335_{(0.073)}$ & $1.104_{(0.024)}$ & \multicolumn{1}{c|}{2.537} & $2.961_{(0.847)}$ & $1.249_{(0.010)}$ & $0.780_{(0.020)}$ & 1.663 \\
  GraphCL~\cite{DBLP:conf/nips/YouCSCWS20}                 & $4.014_{(1.361)}$ & $1.835_{(0.111)}$ & $0.945_{(0.024)}$ & \multicolumn{1}{c|}{2.264} & $3.149_{(0.273)}$ & $1.540_{(0.086)}$ & $0.777_{(0.034)}$ & 1.822 \\
  JOAO~\cite{DBLP:conf/icml/YouCSW21}                    & $3.466_{(1.114)}$ & $1.771_{(0.053)}$ & $0.933_{(0.027)}$ & \multicolumn{1}{c|}{2.056} & $3.950_{(1.202)}$ & $1.220_{(0.028)}$ & $0.710_{(0.031)}$ & 1.960 \\
  GROVER~\cite{DBLP:conf/nips/RongBXX0HH20}                  & $2.991_{(1.052)}$ & $\underline{0.928}_{(0.027)}$ & $0.752_{(0.010)}$ & \multicolumn{1}{c|}{\underline{1.557}} & $2.385_{(1.047)}$ & $0.986_{(0.126)}$ & $\underline{0.625}_{(0.006)}$ & \underline{1.332} \\
  3DInfomax~\cite{stark20213d} & $2.919_{(0.243)}$ & $1.906_{(0.246)}$ & $1.045_{(0.040)}$ & \multicolumn{1}{c|}{1.957} & $2.639_{(0.772)}$ & $\underline{0.891}_{(0.131)}$ & $0.671_{(0.033)}$ & 1.400 \\ 
  GraphMVP~\cite{liu2021pre} & $\underline{2.532}_{(0.247)}$ & $1.937_{(0.147)}$ & $0.990_{(0.024)}$ & \multicolumn{1}{c|}{1.819} & $2.874_{(0.756)}$ & $1.355_{(0.038)}$ & $0.712_{(0.025)}$ & 1.647  \\ 
  KPGT                    & $\textbf{2.314}_{(0.841)}$ & $\textbf{0.848}_{(0.103)}$ & $\textbf{0.656}_{(0.023)}$ & \multicolumn{1}{c|}{\textbf{1.273}} & $\textbf{2.121}_{(0.837)}$ & $\textbf{0.803}_{(0.008)}$ & $\textbf{0.600}_{(0.010)}$ & \textbf{1.175} \\ \hline
  \end{tabular}
\end{table*}
The comprehensive test results suggested the following trends:

\textbf{Observation 1.} 
From the test results, KPGT outperformed baseline methods with considerable margins on all classification and regression datasets under both feature extraction and transfer learning settings. 
For the feature extraction setting, the overall relative improvement was $5\%$ on all datasets, with $3.5\%$ on classification datasets and $8.9\%$ on regression datasets. 
For the transfer learning setting, the overall relative improvement was $3.9\%$ on all datasets, with $2.2\%$ on classification datasets and $8.3\%$ on regression datasets.
This significant improvement demonstrated that KPGT can provide a useful tool for accurate molecular property prediction.

\textbf{Observation 2.} 
Methods incorporating additional knowledge of molecules generally achieved better performance. 
Among the baseline methods, Contextpred+Sup, Masking+Sup and GROVER incorporated additional knowledge of molecules by introducing graph-level pre-training tasks, that is, learning to predict specific properties of molecules, e.g., bio-activities and motifs, using the graph representations. 
Although their node-level pre-training tasks (i.e., predicting the masked nodes or subgraphs) were ill-defined, these methods still generally outperformed other baseline methods that did not exploit additional knowledge, which indicated that incorporating additional knowledge was important to the success of the self-supervised learning methods on molecular graphs.

\textbf{Observation 3.} 
Methods employing high-capacity backbone networks achieved better performance. 
Among all the baseline methods, GROVER also employed a transformer-based model with around 100 million parameters as its backbone network and achieved superior performance in comparison with other baseline methods.
This result further validated our point that a high-capacity backbone network was necessary to yield promising results for self-supervised learning on molecular graphs.

\textbf{Observation 4.}
The methods under the transfer learning setting generally achieved better performance over those under the feature extraction setting.
This result was reasonable because fine-tuning all the parameters enabled the models to capture more task-specific information.
However, on some datasets, KPGT under the transfer learning setting only achieved comparable (i.e., on the SIDER and MetStab datasets) or even worse (i.e., on the BACE dataset) performance compared to that under the feature extraction setting, which may be caused by the catastrophic forgetting problem in the fine-tuning~\cite{kirkpatrick2017overcoming,li2019learn}.
Therefore, exploring different fine-tuning strategies is a potential future direction to further improve the performance of molecular property prediction.
\begin{table}[htpb]
  \centering
  \caption{Comparison results of different backbone networks on classification and regression datasets, measured in terms of AUROC and RMSE, respectively, under the transfer learning setting. 
  }
\begin{tabular}{ccc}
  \toprule
  Backbone            & \makecell{Classification dataset $\uparrow$ \\ \textbf{AVG}} & \makecell{Regression dataset $\downarrow$ \\ \textbf{AVG}} \\ \midrule
  GIN~\cite{DBLP:conf/iclr/XuHLJ19}                 & 0.817                      & 1.229                    \\
  \makecell{Vanilla \\ Transformer} & 0.822                       & 1.260                   \\
  Graphormer~\cite{ying2021do}               & 0.826                       & 1.189    \\
  LiGhT               & 0.829                       & 1.162                   \\ \bottomrule
  \end{tabular}
  \label{table:backbone}
\end{table}
\subsection{Ablation Studies}
\subsubsection{How Powerful is the LiGhT Backbone Network?}
To evaluate the expressive power of LiGhT, we replaced LiGhT in KPGT with other backbone networks, including GIN, vanilla Transformer (i.e., without path encoding and distance encoding) and Graphormer~\cite{ying2021do}. 
For a fair comparison, we restricted all the models to have nearly the same number of parameters (i.e., 3.5 million).
Then we individually pre-trained the models with our self-supervised learning strategy.
As shown in Table~\ref{table:backbone}, our backbone network LiGhT consistently outperformed GIN, vanilla Transformer and Graphormer, which verified that LiGhT can provide a reliable backbone network for self-supervised learning on molecular graphs. 
\subsubsection{How Powerful is the Knowledge-Guided Pre-training Strategy?}
\begin{table}[htpb]
  \centering
  \caption{Comparison results of the pre-trained LiGhT (KPGT) and LiGhT (i.e., without pre-training) on classification and regression datasets in terms of AUROC and RMSE, respectively, under the transfer learning setting. 
  }
  \begin{tabular}{ccc}
    \toprule
     & \makecell{Classification dataset $\uparrow$ \\ \textbf{AVG}} & \makecell{Regression dataset $\downarrow$ \\ \textbf{AVG}} \\ \midrule
    LiGhT                  & 0.803                        & 1.473                   \\
    KPGT                   & 0.843                       & 1.175                   \\ \midrule
    \makecell{Relative\\improvement}                   & $5.1\%$                     & $18.8\%$                 \\ \bottomrule
    \end{tabular}
  
  \label{table:kgp}
\end{table}
\begin{table}[htpb]
    \centering
    \caption{Comparison results between KPGT and baseline methods, which also incorporated the same additional knowledge, on classification and regression datasets, measured in terms of AUROC and RMSE, respectively, under the feature extraction setting.
    }
    \begin{tabular}{ccc}
      \toprule
       & \makecell{Classification dataset $\uparrow$ \\ \textbf{AVG}} & \makecell{Regression dataset $\downarrow$ \\ \textbf{AVG}} \\ \midrule
      MD+FP                  & 0.776                        & 1.698                   \\
      GROVER+FP                   & 0.789                       & 1.541                   \\
      KPGT                   & 0.825                     & 1.273                 \\ \bottomrule
      \end{tabular}
    
    \label{table:semantics}
\end{table}
To evaluate our self-supervised learning strategy, we first compared the prediction performances of the pre-trained LiGhT (KPGT) and the LiGhT without pre-training, both of which followed the same hyper-parameter settings.
As shown in Table~\ref{table:kgp}, the pre-trained model consistently achieved superior prediction performance and the relative improvements were $5.1\%$ and $18.8\%$ on classification and regression tasks, respectively.

Next, we asked whether simply using the same additional knowledge employed in KPGT could provide better results.
We introduced two baseline methods, which also incorporated the same additional knowledge employed in our framework. 
First, we directly concatenated the molecular descriptors and fingerprints employed in our framework as input and then passed through a 2-layer MLP for prediction (denoted by MD+FP).
Note that the previous method GROVER~\cite{DBLP:conf/nips/RongBXX0HH20} had already concatenated the molecular descriptors employed in our framework to the graph representation derived from the readout function.
Here, we further concatenated the fingerprints to such output and passed through a 2-layer MLP for the final prediction (denoted by GROVER+FP).
For a fair comparison, we fixed the parameters of the feature extractors in both GROVER+FP and KPGT (i.e., the feature extraction setting).
As shown in Table~\ref{table:semantics}, KPGT still outperformed the baseline methods.
These results confirmed that our self-supervised learning strategy can potentially learn the abundant semantics of molecules beyond the introduced additional knowledge.
\subsubsection{Effect of Different Masking Rates}
\label{section:masking_rates}
\begin{table}[htpb]
  \centering
  \caption{Ablation results of KPGT with different masking rates on classification and regression datasets, measured in terms of AUROC and RMSE, respectively, under the transfer learning setting.
   }
  \begin{tabular}{ccc}
  \toprule
  Masking rate & \makecell{Classification dataset $\uparrow$ \\ \textbf{AVG}} & \makecell{Regression dataset $\downarrow$ \\ \textbf{AVG}} \\ \midrule
  $15\%$                  & 0.827                        & 1.228                   \\
  $30\%$                   & 0.835                       & 1.216                   \\
  $50\%$                  & 0.843                       & 1.175                   \\ 
  $60\%$                & 0.832                       & 1.186                   \\ \bottomrule
  \end{tabular}
  \label{table:rate}
\end{table}
In our self-supervised learning strategy, we masked a proportion of the nodes and additional knowledge in the molecular line graph.
Here, we evaluated the effect of different masking rates.
In particular, we individually pre-trained KPGT with the masking rates of $15\%$, $30\%$, $50\%$ and $60\%$, and reported the corresponding prediction results on the downstream tasks (see Table~\ref{table:rate}).
The results showed that setting the masking rate to $50\%$ achieved the best prediction performance.
Such an optimal masking rate was much larger than that equipped in the previous methods (i.e., $15\%$)~\cite{DBLP:conf/iclr/HuLGZLPL20,DBLP:conf/nips/RongBXX0HH20}, which indicated that the additional knowledge incorporated in KPGT accurately guided the model to predict the masked nodes and thus enabled it to capture the semantics of molecules. Moreover, recent research in the field of CV also has shown that larger masking rates in the pre-training resulted in better results on the downstream tasks~\cite{he2021masked}.

    \label{sections:experiments}
    
\section{Conclusion} 
    In this paper, we first point out the two major issues of current self-supervised learning methods on molecular graphs, that is, the ill-defined pre-training tasks and the limited model capacity.
Next, we propose a novel self-supervised learning framework, named Knowledge-guided Pre-training of Graph Transformer (KPGT), which consists of two main components, that is, the Line Graph Transformer (LiGhT) and a knowledge-guided pre-training strategy, to alleviate the aforementioned issues and learn more generalizable, transferable and robust representations of molecular graphs to improve the performance of the downstream molecular property prediction tasks.
Extensive test results demonstrated that KPGT can offer superior performance over current state-of-the-art methods on several molecular property prediction datasets.
    \label{sections:conclusion}
\begin{acks}
    \label{sections:acknowledge}
    This work was supported in part by the National Key Research and Development Program of China (2021YFF1201300), the National Natural Science Foundation of China (61872216, T2125007 to JZ, 31900862 to DZ), the Turing AI Institute of Nanjing, and the Tsinghua-Toyota Joint Research Fund.
\end{acks}

\bibliographystyle{ACM-Reference-Format}
\bibliography{sample-base}
\newpage
\appendix
    \section{Appendix}
\subsection{The Ill-Defined Pre-Training Task}
\begin{figure}[htpb]
  \centering
  \includegraphics[width=\linewidth]{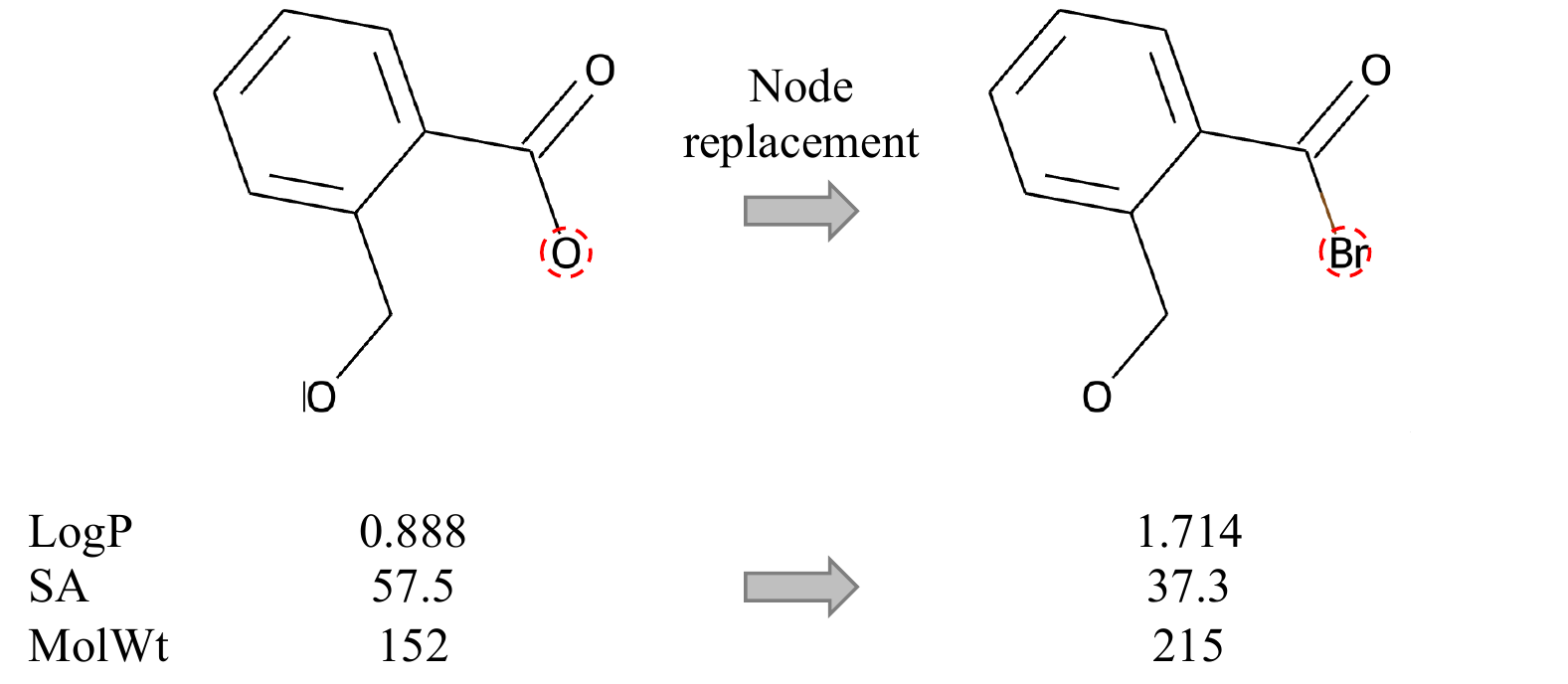}
  \caption{A small modification of a molecule can lead to great changes in its raw properties.}
  \Description{A woman and a girl in white dresses sit in an open car.}
  \label{figure:modification}
\end{figure}

\begin{figure}[htpb]
  \centering
  \includegraphics[width=\linewidth]{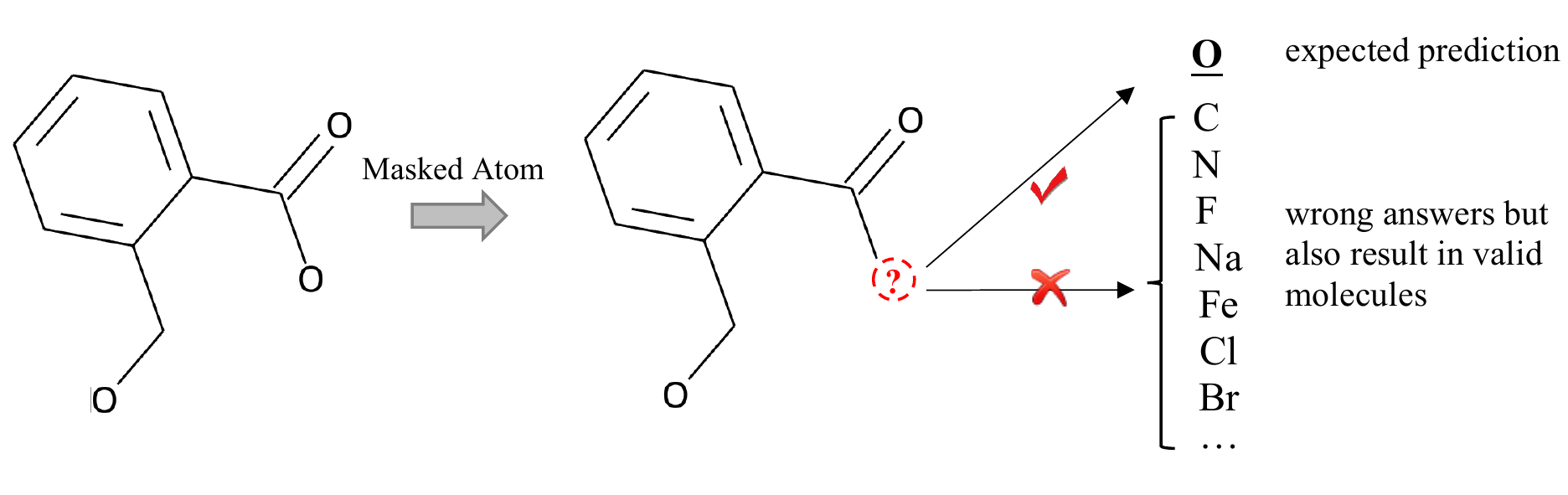}
  \caption{The dependence between the masked node and its neighboring nodes is weak.}
  \label{figure:mask}
\end{figure}
\subsection{Implementation Details}
\subsubsection{Molecular Graph Construction}
\label{appendix:molecule}
\begin{table}[htpb]
  \centering
  \caption{Atom features.}
  \begin{tabular}{ccc}
  \toprule
  Feature & Size & Description \\ \midrule
  atom type                  & 101                       & \tabincell{c}{type of atoms (e.g., C, N, O), by atomic \\number (one-hot)}                 \\
  degree                  & 12                        & \tabincell{c}{number of bonds the atom is involved \\in (one-hot)  }                 \\
  formal charge                  & 1                        & \tabincell{c}{electronic charge assigned to \\the atom (integer) }                  \\
  chiral center                  & 1                        & \tabincell{c}{is charity center (integer)}              \\
  chirality type              & 2                         & R or S (one-hot)                                          \\
  number of H                  & 6                        & \tabincell{c}{number of bonded hydrogen \\atoms (one-hot) }                  \\
  atomic mass                  & 1                        & \tabincell{c}{mass of the atom, divided by \\100 (integer) }                  \\
  aromaticity                  & 1                        & \tabincell{c}{whether this atom is part of an \\aromatic system (integer) }    \\              \\
  radical electrons           & 6                         & number of radical electrons (one-hot) \\
  hybridization                  & 6                        & \tabincell{c}{sp, sp$^2$, sp$^3$, sp$^3$d, sp$^3$d$^2$ \\ or unknown (one-hot) }                  \\ \bottomrule

  \end{tabular}
  
  \label{table:atom_feature}
\end{table}
\begin{table}[htpb]
  \centering
  \caption{Bond features.}
  \begin{tabular}{ccc}
  \toprule
  Feature & Size & Description \\ \midrule
  bond type                  & 5                        & \tabincell{c}{type of bonds (i.e., single, double, triple, \\aromatic or unknown) (one-hot)}                   \\
  stereo                  & 7                        & \tabincell{c}{StereoNone, StereoAny, StereoZ, StereoE, \\StereoCis, StereoTrans or unknown(one-hot)}                   \\
  in ring                  & 1                        & \tabincell{c}{whether the bond is part of a ring\\ (integer)}                   \\
  conjugated                  & 1                        & \tabincell{c}{whether the bond is conjugated \\(integer)}                    \\ \bottomrule
  \end{tabular}
  \label{table:bond_feature}
\end{table}
We abstracted the molecular graph and initialized the features of nodes and edges in the molecular graphs via RDKit~\cite{greg_landrum_2021_5589557}.
Tables~\ref{table:atom_feature} and \ref{table:bond_feature}summarize the atom and bond features used in our experiments, respectively.

\subsubsection{Training Details}
\label{appendix:train}
\begin{table}[htpb]
  \centering
  \caption{The pre-training and fine-tuning hyper-parameters for KPGT.}
  \begin{tabular}{ccc}
  \toprule
  Hyper-parameter & Pre-training & Fine-tuning \\ \midrule
  hidden size & 768 & 768 \\
  number of layers & 12 & 12 \\
  number of attention heads & 12 & 12 \\
  dropout rate & 0.1 & [0, 0.05, 0.1, 0.2] \\
  batch size & 1024 & 32 \\
  learning rate & $2e^{-4}$ & [$1e^{-6}$,$3e^{-6}$,$1e^{-5}$,$3e^{-5}$] \\
  learning rate decay & polynomial & polynomial \\
  weight decay & $1e^{-6}$ & [$0$,$1e^{-6}$,$1e^{-4}$] \\
  Gradient clipping & 5.0 & 5.0 \\
  Adam $\beta_1$ & 0.9 & 0.9 \\
  Adam $\beta_2$ & 0.999 & 0.999 \\
  predictor hidden size & NA & 256 \\
  predictor layer & NA & 2 \\
  mask rate & 0.5 & NA \\ \bottomrule
  \end{tabular}
  \label{table:hyperparam}
\end{table}
All models were implemented in PyTorch~\cite{paszke2017automatic} version 1.10.0 and DGL~\cite{wang2019dgl} version 0.7.2 with CUDA version 11.3 and Python 3.7.

The pre-training and fine-tuning hyper-parameters for KPGT are summarized in Table~\ref{table:hyperparam}.
For KPGT and baseline methods, we tried 48 hyper-parameter combinations of different learning rates, dropout rates, and weight decay listed in Table~\ref{table:hyperparam} via grid search to find the best results for each dataset.

\subsection{Datasets}
\label{appendix:dataset}
\begin{table}[htpb]
  \centering
  \caption{Details of datasets.}
  \begin{tabular}{cccc}
  \toprule
  Dataset & Type & \#Tasks & \#Molecules \\ \midrule
  BACE & Classification & 1 & 1513 \\
  BBBP & Classification & 1 & 2039 \\
  ClinTox & Classification & 2 & 1478 \\
  SIDER & Classification & 27 & 1427 \\
  ToxCast & Classification & 617 & 8575 \\
  Tox21 & Classification & 12 & 7831 \\
  Estrogen & Classification & 2 & 3122 \\
  MetStab & Classification & 2 & 2267 \\
  FreeSolv & Regression & 1 & 642 \\
  ESOL & Regression & 1 & 1128 \\
  Lipophilicity & Regression & 1 & 4200 \\ \bottomrule
  \end{tabular}
  \label{table:dataset}
\end{table}
The detailed information of the molecular property datasets used in our experiments is summarized in Table~\ref{table:dataset}. 
The details of each molecular property dataset are listed below:
\begin{itemize}
  \item BACE is a collection of molecules that can act as the inhibitors of human $\beta$-secretase 1 (BACE-1)~\cite{subramanian2016computational}.
  \item BBBP is a dataset that records whether a molecule can penetrate the blood-brain barrier~\cite{martins2012bayesian}.
  \item ClinTox is a collection of drugs approved through the Food and Drug Administration (FDA) and eliminated due to the toxicity during clinical trials~\cite{gayvert2016data}.
  \item SIDER records the adverse drug reactions of marked drugs~\cite{kuhn2016sider}.
  \item ToxCast contains multiple toxicity labels for molecules obtained through high-throughput screening tests~\cite{richard2016toxcast}.
  \item Tox21 is a public database containing the toxicity of compounds~\cite{tox21}.
  \item Estrogen contains the molecules with known activities towards the estrogen receptors extracted from the ChEMBL dataset~\cite{gaulton2012chembl}.
  \item MetStab is a dataset measuring the half-life time of molecules within an organism~\cite{podlewska2018metstabon}.
  \item ESOL is a dataset recording the solubility of compounds~\cite{delaney2004esol}.
  \item Lipophilicity is a dataset measuring the molecular membrane permeability and solubility~\cite{gaulton2012chembl}.
  \item FreeSolv contains the hydration free energy of small molecules in water from both experiments and alchemical free energy calculation~\cite{mobley2014freesolv}.
\end{itemize}

    \label{sections:appendix}

\end{document}